\documentclass[journal,comsoc]{IEEEtran}
\usepackage[T1]{fontenc}% optional T1 font encoding

\ifCLASSINFOpdf
   \usepackage[pdftex]{graphicx}
   \graphicspath{{./img/}}
   \DeclareGraphicsExtensions{.pdf,.jpeg,.png}
\else
\fi
\usepackage{amsmath}
\usepackage[cmintegrals]{newtxmath}
\usepackage{algorithm}
\usepackage{algpseudocode}

\usepackage{listings}
\makeatletter
\renewcommand{\ALG@name}{Listing}
\makeatother

\begin{document}
\title{Moving Smart Contracts -- A Privacy Preserving Method for Off-Chain Data Trust}%

\author{Simon Tschirner,
        Shashank Shekher Tripathi,
        Mathias Röper,
        Markus M. Becker
        and~Volker~Skwarek %,~\IEEEmembership{Life~Fellow,~IEEE} % <-this % stops a space
\thanks{Simon, Shashank, Mathias and Volker are with the Research and Transfer Centre - Digital Business Processing, Faculty of Life Sciences, Hamburg University of Applied Sciences, Ulmenliet 20, 21033 Hamburg, Germany. Corresponding author's e-mail: shashank.tripathi@haw-hamburg.de.}%
% <-this % stops a space
\thanks{Markus is with the Leibniz Institute for Plasma Science and Technology (INP), Felix-Hausdorff-Str. 2, 17489 Greifswald, Germany.}
}

\maketitle

\begin{abstract}
Blockchains provide environments where parties can interact transparently and securely peer-to-peer without needing a trusted third party. Parties can trust the integrity and correctness of transactions and the verifiable execution of of binary code on the blockchain (smart contracts) inside the system. Including information from outside of the blockchain remains challenging. A challenge is data privacy. In a public system, shared data becomes public and, coming from a single source, often lacks credibility. A private system gives the parties control over their data and sources but trades in positive aspects as transparency. Often, not the data itself is the most critical information but the result of a computation performed on it. 

An example is research data certification. To keep data private but still prove data provenance, researchers can store a hash value of that data on the blockchain. This hash value is either calculated locally on private data without the chance for validation or is calculated on the blockchain, meaning that data must be published and stored on the blockchain---a problem of the overall data amount stored on and distributed with the ledger.

A system we called \emph{moving smart contracts} bypasses this problem: Data remain local, but trusted nodes can access them and execute trusted smart contract code stored on the blockchain. This method avoids the system-wide distribution of research data and makes it accessible and verifiable with trusted software.
\end{abstract}

% Note that keywords are not normally used for peer-review papers.

% potential title: Moving smart contracts - a way to offchain data trust!

% \fi
\IEEEpeerreviewmaketitle

\section{Introduction}


%\IEEEpubidadjcol

\IEEEPARstart{T}{wo} main features of distributed ledger or blockchain technology are data decentralization and immutability. These features lead to trust and transparency in the system. Public permissionless systems ensure most transparency, as they allow anyone to join the network, receive a copy of the stored information and participate in the consensus---the process of deciding what transactions to be added to the ledger. Naturally, the architecture of blockchains mandates the accessibility of the data generally to be public \cite{Xu.42017}. Therefore, this architecture is infeasible in scenarios dealing with data that has to be (temporarily) private such as sensitive data or intellectual property. The development private distributed ledgers addresses this problem \cite{BernalBernabe.2019} (e.\,g., Hyperledger Fabric \cite{LinuxFundation.2020}). However, private platforms reduce the beneficial core properties of blockchains: decentralization and transparency \cite{Kulhari.2018}, while consortia platforms usually still allow public data access.  

A distributed ledger creates a closed environment, allowing participants to trust information and computations by programs stored and executed within it (so-called \emph{smart contracts}). However, this separate environment poses another challenge: The interaction between a blockchain and the outside world, which is usually conducted by distributed ledger technology (DLT) oracles \cite{Xu.42017}. Oracles are services updating a ledger with off-chain data \cite{dltdefinitions}. However, data from outside a blockchain (off-chain data) cannot have the same level of trust as data originated from within it. Furthermore, DLT oracles fail to handle situations where data is intellectual property or too large to be written on a blockchain. In many cases, the result of a computation based on off-chain data is as significant as the data itself. A way to make computation trustworthy is to conduct it on-chain. However, on-chain execution costs become a challenge because each block building node (miner) has to execute the smart contract computing the introduced data. In large networks, this is resource-consuming.

% Therefore, complex computations and computations based on large or sensitive data are infeasible to be executed within a blockchain (network).  

This paper provides a solution to the described problem of introducing a trusted state to a blockchain network based on sensitive and/or big off-chain data. The proposed concept is called \emph{moving smart contracts}. Moving smart contracts provide trusted (reviewed) software, controlled by the blockchain, to a user on request. Furthermore, the network includes so-called \emph{notaries}. A notary is a trusted and authorized full-node that offers off-chain execution using on-chain stored trusted software as a service.  Therefore, a notary can validate any result by repeating a previously conducted computation using the same software and data. If a user wants an on-chain state to be flagged as validated (and therefore to be trustworthy), the raw data needs only to be made accessible to a notary. Furthermore, since the software for off-chain computation is provided by the blockchain, all actors can rely in its integrity and are able to reproduce results.

The primary research objective is to answer the following \textbf{research question}:
\emph{What process must be defined to make a result based on potentially private and large off-chain data to be trustworthy within a blockchain network?}

% The primary research objective is to explore a way to utilize moving smart contracts as trusted software to introduce a validated state based on off-chain data onto the blockchain (without publicly sharing the data). In other words, to develop a secure method of public verification of private data.

Certification of research data is a specific example, requiring complex computations and data privacy. A certificate usually includes storing a hash value of the original research data on a blockchain \cite{Tschirner2021FosteringOD}.
On one hand, research data can involve intellectual property that should remain private before formal publication. On the other hand, this data can include vast amounts of data, meaning that hash value calculation is resource-consuming. This example solves all three requirements for the application of moving smart contracts. Data shall remain private, the computation can be very resource-consuming, and the computation result is as valuable as the input data. Therefore, this example is the perfect use case to evaluate moving smart contracts.  

The secondary objective is thus to implement the use case of research data certification using moving smart contracts.

The remainder of this paper is structured as follows. After discussing the related work and background in section \ref{sec:background}, a theoretical model describes the concept in section \ref{sec:model} in detail. A detailed description of the implementation of academic data sharing in section \ref{sec:implementation} demonstrates the practical feasibility. Before the conclusion (section \ref{sec:conclusion}), in section \ref{sec:evaluation} an evaluation explores the degree to which moving smart contracts solve the use case and their limitations. 

%Research question: 

%(How) Is it possible to push code from the blockchain into a controlled environment, where it is securely executed? 

%Step 1:Proof-of-existence (by the data provider)
%Step 2: verification of integrity by notary
%Step 3: publication, encrypted exchange and withdrawl (= privacy)

%conflicts with blockchain:
%- publish and verify private things
%- revoke information once published
%- create a trusted state space for offchain data

%Feasibility - of the idea
%Full-filling - its purpose

%Three possible points:
%1. Introduction of validated state based on off-chain data
%2. Preserving data privacy
%(3.Saving calculation resources)

%Comment from Mathias: In order to make real world information available inside the blockchain, oracles are used to provide such data. But to just transfer/inject data is frequently not enough, because in addition, computation is often needed. This adds another element to the process: the code used for computation – which itself requires a level of trust. MSC adds trust to the computation which can be pushed to the edge of the network (notary nodes).
%Comment from Shashank: ...or, securing the privately possessed data on publicly accessible platforms.

\section{Background}
\label{sec:background}

%The typical restrictions of blockchain environments, e.g., constraints on on-chain data storage or processing and the need to include potentially insecure, off-chain data into a blockchain environment, have led to numerous developed principles and design patterns to circumvent these restrictions. 

This section briefly presents the state-of-the-art from a theoretical point of view and discusses, how moving smart contracts differ from other approaches.

%  Following paragraph should be removed completely?
A mechanism build solely for local machines face many kinds of threats like attacks on control flow of program, masquerading malicious code, malevolent writes to memory or run-time attack. A hostile unit can inject its own program that will have all the permissions as the intended original program \cite{One1996}. Therefore, adding checks to the application is not sufficient to counter such attacks. A weakly written program is also susceptible to attacks. A common example includes buffer overflows \cite{Foster2005} that allows the adversary to modify arbitrary memory locations for its benefit. With the help of blockchain, moving smart contracts addresses these threats. 

\subsection{Blockchains and Smart Contracts}

A blockchain is a list of blocks containing transaction data linked together with cryptographic hashes, constructing a Merkle tree, distributed over several network peers (nodes).
Any change in a transaction would change the Merkle tree's root address, eventually changing the address of a block. Hence, the block is invalidated. With proper implementation of consensus algorithms like Proof of Work, \cite{SatoshiNakamoto.15052021}, Proof of Stake \cite{ethereum}, or Practical Byzantine Fault Tolerance \cite{MiguelCastro.1999}, a blockchain-based peer-to-peer network can become tamper-proof.

% Haber and Stornetta \cite{Haber.1991} presented the concept of a cryptographically secured chain of blocks for digital timestamping of modifiable media like text, audio, or video. Later, they upgraded their solution \cite{Bayer.1993} by incorporating Merkle trees \cite{Merkle.1989} which enabled clustering more documents inside a single block.

% Bit Gold proposed by Sabzo \cite{NickSzabo} marked one of the earliest attempts of creating a digital currency, but it was never implemented. The first successfully implemented milestone of the blockchain concept was achieved with Bitcoin \cite{SatoshiNakamoto.15052021}. It is a peer-to-peer transaction-based digital currency. Although Bitcoin removes the need for a central authority in a distributed environment, the application is almost limited to digital currency transfer. 

Buterin proposed the idea of the Ethereum blockchain that can perform computations besides being a peer-to-peer network \cite{ethereum}. \emph{Smart contracts} are small programs, giving rules for performing computations. Szabo first proposed the term smart contract in 1994 to refer to self-executing codes  \cite{NickSzabo}. In Ethereum, smart contracts are self-executing and tamper-resistant programs deployed on the blockchain. Since the execution of smart contracts happens on-chain, it utilizes the nodes' resources. All consensus-creating nodes of the respective blockchain network carry out the smart contract execution in parallel; the resulting redundancy leads to high computational and storage expenses. Therefore, it is common for an application to use off-chain resources and oracles \cite{X.Liu.2018}, e.g., for computationally extensive processing and large storage requirements.

In centralized systems, a central authority is responsible for creating a valid state based on given transactions. Furthermore, the users depend on the central authority to maintain their assets, like money in the case of banks. Blockchains, on the other hand, provide a system where participating peers make decisions in the network reaching a consensus following a consensus algorithm \cite{consensusAlgo}. In terms of permission required for the participation in consensus-creating procedures, blockchains can be classified into two major types:

\paragraph{Permissioned (consortia platforms)}
Only authorized nodes are permitted to participate in consensus-creating procedures.
Depending on the blockchain's governance design, nodes can, for example, become authorized through voting procedures. In some cases, there are different roles assigned to nodes in hierarchical order. The most common examples include Ripple and Eris.

\paragraph{Permissionless}
Permissionless blockchains are publicly accessible. Any user can join the network, become a node, perform transactions, participate in creating consensus and reading the transactions conducted by other network participants \cite{blockchainTaxonomy}. The typical examples of permissionless blockchains are Bitcoin and Ethereum. The advantages of such blockchains are a high degree of decentralization, transparency, and availability open source.

\subsection{Adding External Information Into Blockchain Environments}

An oracle is an agent that facilitates the interaction of a blockchain with the external world \cite{SinKuangLoXiweiXuMarkStaplesLinaYao.,XiweiXuandothers.2016}. Usually, it is a trusted third party in the system, which could be software or a human feeding the data manually. Town Crier \cite{TownCrier} and Corda \cite{Corda} have a single oracle responsible for data transfer. Augur \cite{Augur}, Gnosis, MS Bletchley \cite{gnosis_MS_bletchley} and ChainLink \cite{chainlink} propose to have multiple (independent) oracles. Multiple oracles vote for the correct answers mitigating the possibility of a single-point-of-failure. Although oracles can address off-chain communication,  data integrity remains an issue. There is a risk that data from oracles could be manipulated or bogus; therefore, the smart contract at the receiving end must build a mechanism for data check \cite{HamdaAlBreiki} but there is always a risk at the intersection of on and off chain. Also, oracles work in an external environment; consequently, undesired events can obstruct the operation of the whole system. With the concept of moving smart contracts, data/results written to the blockchain can be validated by notaries. Errors in the off-chain execution of software are made visible since the computation is conducted in the process of validation by two independent parties, using the same trusted software.

% With the concept of moving smart contracts, data outside the blockchain is certified by moving code from inside the blockchain to its edge for verification of data.

Blockchain technology, especially public blockchains, has on-chain resource limitations making data storage infeasible in case of enormous data size. Therefore, researchers have tried to solve this problem by using off-chain storage. Eberhardt and Tai \cite{JacobEberhardt.2017} have discussed how off-chain computation and storage can help resolve the on-chain issues. They have considered five applications where off-chain processing can address on-chain limitations. They have discussed a method to demonstrate how a large amount of data can be associated with a smart contract. The reference to data stored in an addressable content system is stored in the blockchain. The reference is the hash of the data. Storing the hash as a variable of the smart contract ensures that a change of the data is detectable. Although, in this method, the addressable content system has to be trusted for hosting the data. The other possible solution includes decentralized storage systems, such as the InterPlanetary File System (IPFS) \cite{IPFS-Benet.2014}. However, IPFS has no implementation of intellectual property management \cite{Ito.2019}. On a technical level, the storage for big data files on a distributed file system is not feasible as it requires resources from multiple network participants. IPFS stores parts of a file at multiple locations causing redundancy. Thus, ensuring data availability even when a node in the network goes offline. There are some projects that explore the possibilities to ensure correct execution, e.g., like Ethernity \cite{ethernity-patent}, via trusted environments, this paper aims to choose a more straightforward approach discussed in section \ref{sec:model} and \ref{sec:implementation}.

% Perun (secure state channels) – mostly financial 

% PoX (proof of execution), – similar problems, however now description of their approach – patent 
% Ethernity \cite{ethernity-patent}

%The patent application of Ethernity\cite{ethernity-patent}, proposes a solution for secure computing of data with privacy control. The patent does not openly describe the implementation details but the claims point toward using a secure environment for computation to guarantee complete privacy. Their objective and implementation is noticeably different from the moving smart contracts. While they focus on secure environment, moving smart contracts focus on security of data. Moving smart contracts do not allow custom software for data processing, rather the software is manually examined and stored on-chain. 

% - custom code
% - technical solution seems sophisticated, however, published information not with technical details. Focus on security/trusted environments, hardware solutions
\subsection{Use Case -- Certification of Research Data}
%Status: Ready to review, add reference

Sharing of research data is an important aspect in open science \cite{openScience} and sometimes even required by research funds such as by ``Guidelines for Safeguarding Good Research Practice'' of the German Research Foundation (DFG\footnote{https://www.dfg.de/en/research\_funding/principles\_dfg\_funding last \mbox{access} June 7, 2021}). Especially in data-driven science, provenance (authorship and integrity) and reliability of data and metadata have to be proven. Therefore, a neutral and immutable platform that researchers can use to certify their data brings significant benefits to the research community \cite{DIONE}. The certification consists of creating a hash (serving as a fingerprint) of the research data, and storing this hash together with a trusted timestamp on the blockchain. At this stage, the certification proves the existence of the data at a particular time. Including further references to the researcher/data origin to the on-chain certificate, provenance can also be proven, even after publication.

%To have a concrete reference, we developed the moving smart contracts with a use case in mind, coming from the QPTDat research project. QPTDat aims to foster sharing and reusing of research data \cite{DIONE}. One aspect is the certification of data provenance by placing a hash value of the research data on the blockchain; another aspect is the quality control of the related metadata sets.
One very common method for such an proof-of-existence is publishing a data-certificate by their cryptographic hash value. Later, this data can be verified by its hash value as it is sufficiently pseudo-random and unique.
% Here, the primary aspect is data privacy. 

%Completeness and soundness of metadata is an important quality aspect. A system based on an ontology generates rules that have to be evaluated to ensure metadata quality. As common smart contract executions on blockchains are costly, the needed calculation resources would be expensive. Therefore it is desired to execute rule evaluation only on one blockchain node instead of the whole network. Here, the primary aspect is avoidance of calculation costs. 
As the only issue, the hash computation with a local algorithm can be faked. The blockchain itself can only check the validity of the hash, if this data is stored on the blockchain. But this again shall be avoided due to its size. Consequently, moving smart contracts provide a compromise: allowing initial data privacy and later memory resources without trading in trust entirely. 

\subsection{Bloxberg Blockchain Network}
%Status: Ready to review

The Bloxberg consortium was founded in 2019 with the aim of providing a blockchain for science, i.e. to establish an alternative infrastructure for academic purposes \cite{Bloxberg-Whitepaper}.
At the time of writing this paper, the consortium consists of around 40 academic institutions globally. Collectively, they are providing the Bloxberg blockchain\footnote{https://bloxberg.org last access May 14, 2021}.

Bloxberg is a consortial (public permissioned) blockchain network based on Ethereum. Only academic institutions can be voted into the consortium; members are automatically entitled (and required) to run a full node (participating in creating the consensus). As consensus protocol, the Proof-of-Authority algorithm Aura\footnote{https://openethereum.github.io/Aura last access May 14, 2021} is used.

The academic orientation of Bloxberg fits well to the use case of research data certification, the primary use case to evaluate the moving smart contracts in this paper. Additionally, the author's institution (HAW Hamburg) is a member of the Bloxberg consortium, which allows deploying and testing the server-side tool that is part of the suggested implementation. Thus, Bloxberg is the natural choice for the exploration of moving smart contracts.
\section{Model}
\label{sec:model}

This section describes the idea of the moving smart contracts in detail. It outlines the theoretical process, how parties in the process can be assured to use the same software to achieve a result based on off-chain data and how the moving smart contract evaluates that the result is trustworthy. In other words, a concept to introduce a validated state based on off-chain data onto the blockchain.

%The design of moving smart contracts bases on consortial Ethereum blockchains. 
% The argument to use Bloxberg has been made in the background section. We need to argue somewhere, that and why we focus on consortium-based chains. Either here or in the end of the background.

\subsection{General Idea}

Given the usage of Bloxberg, this process is tailored toward public permissioned blockchains. It utilizes the additional portion of trust or stake (since the network needs to approve on a new consortium member) that is connected to a full node of the network. It further aims to transfer the possibility of keeping data private from private blockchains onto public blockchains.

%Describe Actors
There are four kind of actors, interacting with the moving smart contract: \textbf{The client}, this is the party that owns or provides the off-chain data, in the example of research data certification, this would be the researcher, \textbf{the notary} that can act as a validator of the published provenance if requested by the client, \textbf{moving smart contract} that provides and govern the ecosystems and lastly, \textbf{the peer} can request the client's data and verify the results.

\textbf{At first}, only the client obtains or holds the data to be processed. The first important difference when using moving smart contracts is that the software for the processing is available via the blockchain. In the first step, the client downloads that software and processes her data. The result is written back on the blockchain and thus publicly available. 

The result stored on the blockchain can prove data existence at this stage, but it is impossible to guarantee the correct execution of the software stored on the blockchain. However, instead of utilizing complex solutions, moving smart contracts rely on the validation of Alice's execution in later steps.

\textbf{The second} step is an optional step for pre-public validation to increase trust in the data. The basic idea is that the researcher selects a trusted notary in the blockchain network. The researcher can now transfer his data confidentially to Bob, who can validate and confirm or create the PoE-hash. As the notary node is part of the blockchain, it uses and accesses the validated and trusted verification smart contracts on the blockchain.

\textbf{At last, during the third step}, three options exist. 
\begin{itemize}
    \item The researcher can publicly share data by adding its access link to the PoE on the blockchain, and the reference to the written publication including scripts or smart contracts for evaluation. Now everyone can download the data and software stored on the blockchain to validate the results independently.
    \item Alternatively, the researcher can decide to share data privately with selected partners. They will need the same information as before (link to data, software, and result), including, e.~g., credentials to access the data.
    \item As a  last option it is possible to withdraw the result. If for some reason, the researcher decides not to share data or results anymore, he can use a smart contract on the blockchain to mark the previously submitted PoE or even results as invalid. 
\end{itemize}
In summary, the steps to be covered by moving smart contracts are:
\begin{enumerate}
    \item \textbf{Proof of existence (certification)}---the client processes its data and stores the result on the blockchain.
    \item \textbf{Validation}---the client picks a notary to confirm the result from the previous step.
    \item Further steps
    \begin{enumerate}
        \item \textbf{Publication}---client decides to share its data publicly; everyone will be able to access the client's data and to validate the certificate/processing.
        \item \textbf{Private sharing}---client decides to share its data privately, a selected peer gets access to the data protected by credentials; the peer can access the client's data and validate the certificate/processing.
        \item \textbf{Withdrawal}---client withdraws result.
    \end{enumerate}
\end{enumerate}

%\subsection{Connection to the Scenario -- Certification of Research Data}

Having the certificate additionally validated by a notary (the second step above) is optional in this scenario. However, using a notary, the network has strong evidence that the data existed, and Alice rests assured that later validation of the certificate will succeed.

\subsection{Process flow of privacy preserving research data publication and moving smart contracts}

This section gives a more technical perspective on moving smart contracts based on the steps presented above. 

\begin{figure}[!t]
\centering
\includegraphics[width=3in]{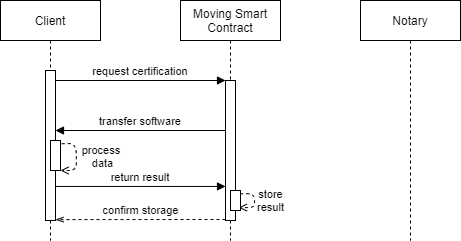}
\caption{Step 1: A client processes its data and stores the result on the blockchain (proof-of-existence).}
\label{fig_s1}
\end{figure}

The process starts with the initial certification or PoE. In figure \ref{fig_s1}, the researcher wants to create a proof of existence for his data. The software needed for this certification is stored on or is accessible via the blockchain. The moving smart contract---an actual smart contract deployed on the blockchain---transfers the software for the certification request to the client. The client processes the data and returns the result to the moving smart contract, which takes care of storing the result on the blockchain. From now on, the client can prove that the data processing happened at a particular time.

The sub-step \emph{process data} takes place in several steps. It means that the active part processes the data with a piece of software. In the typical case, the software will always be the same in one instance of the process, as the moving smart contract provides it. The data should, in any case, be the data that the client initially processed. If not, any later processing will yield a different result.

\begin{figure}[!t]
\centering
\includegraphics[width=3in]{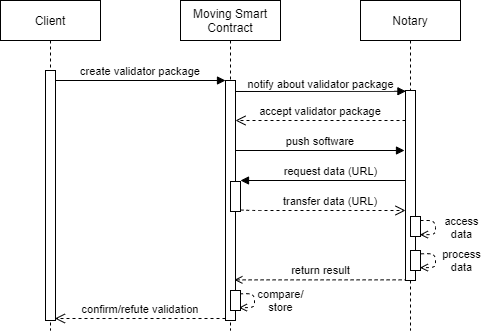}
\caption{Step 2: A client selects a notary to verify its result from the first step.}
\label{fig_s2}
\end{figure}

The previous proof has one problem: As long as no other trusted party confirms it, this self-certification is of limited value. The optional second step (see Figure \ref{fig_s2}) increases trust: A notary within the blockchain network--a node of higher and central trust---gets the task to validate the certification from the previous step. In order to do so, the client sends a \emph{validator package} to the moving smart contract. 

The validator package is a data structure consisting of the following:
\begin{itemize}
    \item recipient---optional, but as long as data privacy is an issue, these packages address a certain receiver, either a peer or a notary,
    \item reference to on-chain data---optional, a reference to the result the client initially stored on the blockchain,
    \item software reference---a reference to the required software that reproduces the processed data from the raw data on the client,
    \item external data reference---typically a URL under which the original data is accessible,
    \item credentials---optional, credentials needed to access the data, in case it is to be kept private.
    %\item Secret -- the moving smart contract will hash the secret: if the result matches the provided secret hash, the smart contract can tie the validator package and the initially stored result together
\end{itemize}

The moving smart contract uses the receiver information to notify, in this case, the notary. The notary can accept the package, and, in consequence, the moving smart contract pushes the software onto the notary. The notary then reads the data, which shall be verified, on a separate channel outside the blockchain from the client. This sub-step is also common to several steps. It includes receiving the location of the data (\emph{request data(URL)}) to process and its retrieval (\emph{access data}), in this case, typically requiring credentials, to retain data privacy.

The notary now processes the data and returns his result to the moving smart contract. In the recurring sub-step \emph{compare/store}, the moving smart contract compares the notary's result with the initially stored client's result. A match yields in confirmation of the certificate, otherwise a refutation of the validation. In the latter case, the client could renew the validation request, as something might have gone wrong, e.g., during data transmission. However, retries should be limited.

\begin{figure}[!t]
\centering
\includegraphics[width=3in]{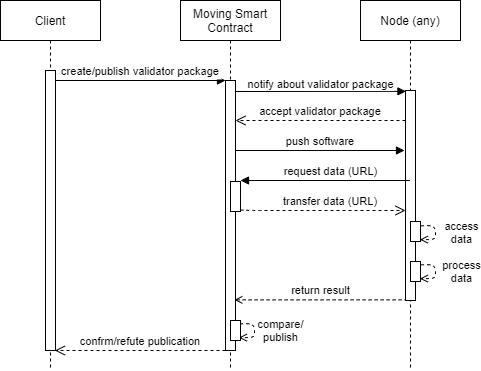}
\caption{Step 3a: A client publishes the original data, similar to step 2, a node from the blockchain network verifies the client's result before the data can be published.}
\label{fig_s3a}
\end{figure}

Eventually, the client might choose to publish her data publicly. Figure \ref{fig_s3a} depicts this step. This step resembles the previous step with a few differences: The validator package does not contain a receiver and is publicly available via the moving smart contract. Consequently, now any node of the blockchain network may execute the validation step previously done by the notary. On successful comparison, the moving smart contract confirms the publication.

\begin{figure}[!t]
\centering
\includegraphics[width=3in]{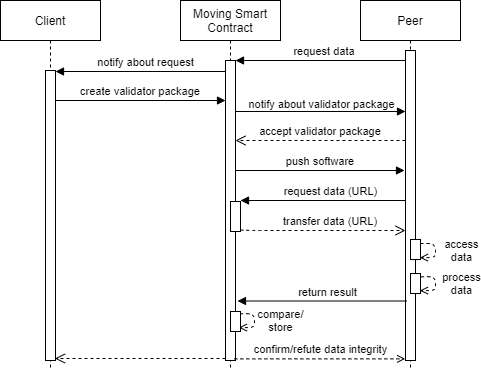}
\caption{Step 3b: A peer request to (privately) receive the client's data, only when the peer confirms the client's result, integrity of the data is proven}
\label{fig_s3b}
\end{figure}

Before publication, a client might also want to share her data privately. Figure \ref{fig_s3b} depicts this step, and again, it resembles the two previously described steps. However, in this step, the peer initializes the process, asking the client for her data. According to the request, the client creates another validator package, this time directed towards the peer. The peer gets notified via the moving smart contract, software and data transfer to and processing on the peer functions similarly. Afterward, the peer writes the result back to the moving smart contract. The moving smart contract compares the new result to the stored result. A match confirms data integrity, and the peer is free to proceed with the transferred data, which can now be regarded as validated.

\begin{figure}[!t]
\centering
\includegraphics[width=3in]{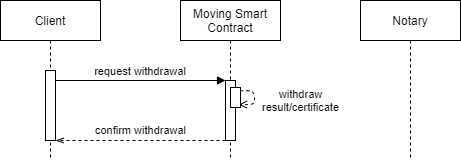}
\caption{Step 3c: Client withdraws the result from first step.}
\label{fig_s3c}
\end{figure}

The final option is for the client to withdraw the data. As depicted in figure \ref{fig_s3c}, this step is relatively simple: The client indicates withdrawal to a smart contract; after that, the smart contract on the blockchain adds a note of withdrawal to the information stored on the blockchain. The result should not be regarded as certified anymore.

%sufficiently described above!?
%\subsection{Jobs} % Validator Package

%Here we could add some details/a drawing of jobs, we could also refer to Ethernity \cite{ethernity-patent} and state differences in our jobs.

\subsection{Software for execution uniformity}

The software is either stored in a transaction on the blockchain or on off-chain storage. In the latter case, design patterns such as \cite{Eberhardt.2017} ensure the integrity of the software, e.g., securing it with a hash stored on the blockchain (cf. \cite{Eberhardt.2017}, Section 4.3). 

The software can exist in the form of source code that needs to be compiled first, in the form of byte code or a container. Each form has advantages and disadvantages. On the one hand, source code is lightweight, but it requires an executable environment installed on the systems supposed to run the software. On the other hand, containers are complete in terms of the environment but are bulkier than source code. Additionally, the software must guarantee the same result independent of the used platform, which is more challenging when compiling source code separately on each target.

At last, the software must be (manually) controlled first. Otherwise, it is possible to introduce malicious code to the notary or any party to execute the software.
\section{Implementation}
\label{sec:implementation}

The goal of the implementation is to prove the feasibility of the presented moving smart contracts. The implementation covers step 1 and 2 from the previous section (see Figure \ref{fig_s1} and \ref{fig_s2}). As discussed, steps 3a and 3b are, to a large extent, similar to step 2. The main difference is that the third party in step 2 is a notary, i.e., an actively selected node of the blockchain network, while it can be any node of the blockchain network in step 3a. In step 3b, the restriction of the third party to be a member of the blockchain network does not apply anymore. Therefore it is safe to assume that the feasibility of implementation of step 1 and 2 implies the feasibility of step 3a and 3b. Due to its minimal extent, the implementation of step 3c is omitted.

\subsection{Implementation Details}
% (Concept of the network)
The prototype system consists of three components: the client-side application, the smart contract deployed to Bloxberg (public permissioned blockchain), and the notary side server application. The client-side and notary (server)-side applications are programmed in the Python programming language. The smart contract is written in Vyper---a Python-like programming language for Ethereum smart contract development. Pseudo-code presents relevant parts of the implementation.

\subsubsection{Client-Side Implementation}

% Code needs to be updated to include secret, description in text needs to be added

Clients are actors, e.g., researchers, who use the provided functionalities of the system. To do so, they use the client-side tool---an application running on their local computer. Since running an own node to interact with the blockchain is not feasible for most clients, a trusted third party node can serve as an entry point for blockchain interaction.

The pseudo-code in listing \ref{algo:algo1} shows the procedure to certify data (cf. Figure \ref{fig_s1}), in the client-side tool, in an abstract way. The communication with the moving smart contract, i.e., the smart contract deployed on Bloxberg implementing the functionality of the moving smart contract, happens via the \verb!sc_interface!. The user is prompted for the data path (e.g., on her local computer). The tool then requests the software from the smart contract, compiles it (\verb!create_executable_script!), and runs it, passing the data from the inserted path onto the executable. In this example, the selected software calculates the hash value on the passed data, the result is written back onto Bloxberg via the smart contract interface.

% This is the code for the client-side tool -> data certification.
\begin{algorithm}
\caption{Client certifies data}
\label{algo:algo1} %for referencing later
\begin{algorithmic}[1]
\Procedure{certify\_data($sc\_interface$)}{}
    \State $s \gets$ false
    \State $code\_id \gets "HASH\_FUNCTION"$
    \State $data\_path \gets$ \textit{user input}
    \State $code \gets sc\_interface.request\_code(code\_id)$
    \State $program \gets create\_executable\_script(code)$
    \State $data\_hash \gets execute(program, data\_path)$
    \State $secret, secret\_hash \gets generate\_secret()$
    \State $s \gets sc\_interface.create\_certificate(data\_hash, $
        \State $\qquad secret\_hash, code\_id)$
    \State \Return $[s, data\_hash, secret]$
\EndProcedure
\end{algorithmic}
\end{algorithm}

To perform the step of data validation via a notary (cf. Figure \ref{fig_s2}), the clients use the tool for creating the requests in the form of validator packages. Each validator package is organized as an object within the dedicated smart contract. Each contains information regarding who shall using what code to process which data.

Listing \ref{algo:algo2} shows the procedure to create validator packages. First, the tool prompts the user for all relevant input: IDs of the notary and software to select and a URL pointing to the data. Not shown in the pseudo-code: Users can also use the tool to list possible notaries and software to choose from. The next step involves the selection of the notary node of interest. In principle, the selected notaries are the only party in the blockchain network to access the clients' data. However, if data and its URL are provided unencrypted, anyone can read the data, as the access to the information on Bloxberg is public. If data should remain private, which typically is the case for this step, users should select to use encryption. First, the data storage requires credentials (\verb!data_pw!) for data transfer/decryption. Second, either the credentials or additionally the provided URL need to be encrypted. If the data is to be shared with a node of the blockchain network (which is part of the definition of a notary), the tool uses that node's public key for encryption, which is available through the blockchain network, accessed via the smart contract interface.

%Therefore, the client selects a trusted notary, possibly his/her home institution's node, that will be responsible for the computation of the selected code on the given data for the process of validation. 

% This is the code for the client-side tool -> data validation.
\begin{algorithm}
\caption{Client initiates data validation by notary}
\label{algo:algo2}
\begin{algorithmic}[1]
\Procedure{initiate\_data\_validation($sc\_interface,$}{}
        \State $\qquad data\_hash, secret)$
    \State $vp \gets$ false
    \State $notary\_id \gets$ \textit{user input}
    \State $software\_id \gets$ \textit{user input}
    \State $data\_url \gets$ \textit{user input}
    \State $data\_pw \gets$ \textit{user input}
    \State $encrypted \gets$ \textit{user input}    
    \If{$encrypted =$ true}
        \State $pub\_key \gets sc\_interface.get\_pkey(notary\_id)$
        \State $data\_url \gets encrypt(data\_url, pub\_key)$
        \State $data\_pw \gets encrypt(data\_pw, pub\_key)$
    \EndIf
    \State $vp \gets sc\_interface.create\_vp(data\_hash, $
    \State $\qquad notary\_id, data\_url, data\_pw,encrypted)$
    \State \Return $vp$
\EndProcedure
\end{algorithmic}
\end{algorithm}

%After selecting code and notary-node, the client provides the URL of his data. If a password secures the data access, the client can also provide it in this step. The client will have the option to encrypt the URL and password (if provided), so it is not publicly accessible. In the case of encryption, the public key of the selected notary node is used to encrypt the information for data access. 

After the tool has collected all information, the validator package is created via the smart contract interface. The process continues on the notary.

%As the next step, it uses the code to hash the data. In the context of the given use case, the code includes the SHA256 hashing algorithm, but it can be any other hashing algorithm. The client-side tool accesses the data and passes it to the code. After successfully executing the code, the job request object is published on the blockchain – providing the selected notary with the necessary information in real-time. At the same time, the client-side hash is published on the blockchain, which the notary will later validate.

\subsubsection{Notary-Side Implementation}

%Ready to review

% The selection of notary nodes is done through consensus between all the present notary nodes. When an entity wishes to be part of the network as a notary node, it applies to it by running a node on its server and applying to the network. Then, all the notary nodes vote for its inclusion. After successful voting, the organisation becomes the new notary node.

A notary is a supervisory node and, in this case, part of the Bloxberg network, i.e., a public or private research institute, university, or organization. A node needs to run the program shown in Listing \ref{algo:algo3} to become a possible notary and register to the moving smart contract (that functionality is available but omitted in this paper to focus on the primary functionality). 

\begin{algorithm}
\caption{Receive and process validator package}
\label{algo:algo3}
\begin{algorithmic}[1]
\Procedure{validate\_data($sc\_interface,$ $notary\_id$)}{}
    \State $KEY \gets self.PRIVATE\_KEY$
    \While{true}
        \State \textit{vp} $\gets sc\_interface.get\_vp(notary\_id)$
        \State $data\_hash \gets vp["data\_hash"]$
        \State $data\_url \gets vp["data\_url"]$
        \State $data\_pw \gets vp["data\_pw"]$
        \If{$vp["encrypted"] =$ true}
            \State $data\_url \gets decrypt(data\_url, KEY)$
            \State $data\_pw \gets decrypt(data\_pw, KEY)$
        \EndIf
        \State $code \gets sc\_interface.accept(vp)$
        \State $program \gets create\_executable\_script(code)$
        \State $data \gets retrieve\_data(data\_url, data\_pw)$
        \State $n\_hash \gets execute(program, data)$
        \State $sc\_interface.send\_notary\_result($
        \State $\qquad vp, n\_hash)$
    \EndWhile
\EndProcedure
\end{algorithmic}
\end{algorithm}

Notaries continuously stay in listening mode. A notary triggers when a client publishes a validator package on the blockchain directed to the notary itself. The triggered notary extracts information from the validator package lying on the blockchain. On acceptance, the moving smart contract transfers the software to the notary. If the client encrypted data access, the notary uses its private key to decrypt the information. The notary retrieves the data from the URL in the validator package and processes it with the software it received from the moving smart contract. After execution, the notary sends the validator package alongside its result back to the moving smart contract; the notary cleared the validation package.
% -------------

\subsubsection{Moving Smart Contract Implementation}

% Todo: Insert class diagram --> done

% Real code from sc.vy
%\lstinputlisting[language=Python, basicstyle=\tiny, label="code:sm", caption=Smart Contract Code]{sc.vy}

%\subsubsection{Purpose of Smart contract}
%Smart contract is the core of the whole architecture. It defines the functionalities that are callable by the client-side tool as well as notary node. The smart contract provides the following functionalities:
%\begin{itemize}
%    \item Enforces schema to the addition of new notary nodes, codes or creation of a validator package.
%    \item Creating a new validator package.
%    \item Registering a notary node to the network.
%    \item Adding a new code to the list of codes for hashing the data.
%    \item Publishing the hashed results to the blockchain.
%    \item Fetching the validation request by validator package id.
%    \item Fetch the meta data of notary by notary id.
%    \item validate the results by matching the hashed results by the notary and the client.
%\end{itemize}

This subsection describes the implementation of the moving smart contract, which is deployed on the Bloxberg blockchain. The moving smart contract is the main communication medium for clients and notaries; it brings together the previous two subsections. It also keeps track of the available software, the available notary nodes, created and published results, and the publication of validator packages. In the context of the underlying use case, the software represents solely hashing algorithms written in Python. The prototype includes the algorithms of SHA-2 (SHA-256)  and SHA-3 (Keccak). Consequently, the execution of the code results in hashes that can be interpreted as certificates and prove the existence of the research data. These are the results stored on Bloxberg. 

\begin{figure}[!t]
\centering
\includegraphics[width=3.4in]{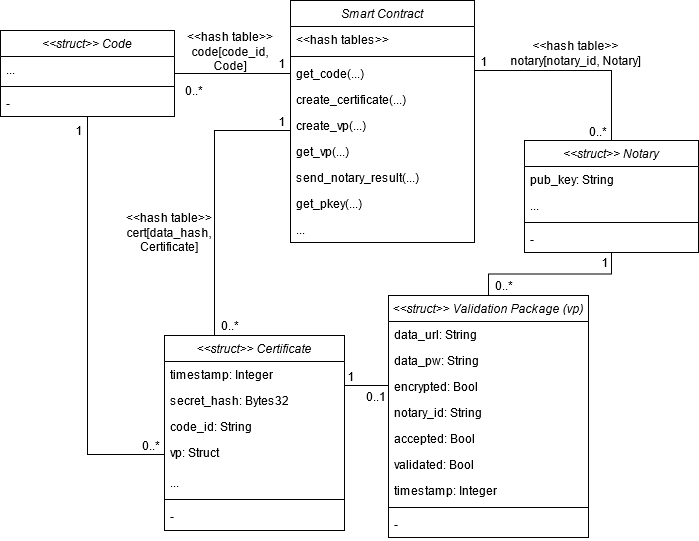}
\caption{Architecture of the smart contract}
\label{fig_class_diagram}
\end{figure}

The moving smart contract defines a couple of structures (structs), depicted in Figure \ref{fig_class_diagram}.

Both notary nodes and software are organized as objects and stored within callable and extendable registers (hash tables). Since the content lies on the blockchain, an append-only data structure, the content added to the hash tables cannot be altered afterward. It applies to the stored software as well. Therefore, and because it will be executed on blockchain nodes, the software must be peer-reviewed and able to execute or compile. Additionally, large software should be stored off-chain, secured with an on-chain hash-value for verification. The implemented hashing algorithms used in the example are small enough to be stored on-chain. They can be retrieved via the \verb!get_code! procedure. This procedure returns the elements of the code hash table stored under the given name.

%Pseudo-code for smart contract functionality
%\begin{algorithm}
%\caption{Get code from the blockchain (smart contract functionality)}
%\label{algo:sc:code}
%\begin{algorithmic}[1]
%\Procedure{get\_code($code\_id$)}{}
%    \State \# Self is referring to the contract instance
%    \State \Return $self.code[code\_id]$
%\EndProcedure
%\end{algorithmic}
%\end{algorithm}

The moving smart contract adds new execution results, in this case certificates, in the \verb!cert! hash table. This procedure is shown in Listing \ref{algo:sc:certify}. The proposed implementation takes advantage of a requirement from the use case: The result is always a hash value of the research data that is unique and can be used as a key for the hash table. Other use cases would require the creation of a separate hash key. The first step is to assert the novelty of the result to add. If this assertion fails, the research data has already been certified before; another certification is not allowed. Otherwise, the certification is added alongside a timestamp and a hashed secret. The hashed secret is an additional way to prove ownership: only the client originally requesting certification knows the secret. To initiate further steps in the process of certification (depicted in Figure \ref{fig_s2} to \ref{fig_s3c}), a client would need to reveal this secret to the moving smart contract. This prevents third parties from initiation of further steps.

%Pseudo-code for smart contract functionality
\begin{algorithm}
\caption{Certify data (smart contract functionality)}
\label{algo:sc:certify}
\begin{algorithmic}[1]
\Procedure{send\_data\_hash($data\_hash, code\_id,$}{}
        \State $\qquad secret)$
    \State \# Abort execution if data is already certified
    \State \textbf{assert} $self.new\_entry(data\_hash)$
    \State \# Create certificate on-chain
    \State $self.cert[data\_hash].timestamp \gets self.time()$
    \State $self.cert[data\_hash].secret\_hash \gets secret$
    \State $self.cert[data\_hash].code\_id \gets code\_id$
    \State \Return true
\EndProcedure
\end{algorithmic}
\end{algorithm}

Listing \ref{algo:req_validation} is the counterpart to the client's validation request Listing \ref{algo:algo2}. It starts with two assertions to ensure that the data to validate exists and the initiation request includes the correct secret. Since the secret is now revealed, it is replaced with another secret to authenticate future access to that certificate. If everything is correct, the moving smart contract creates a validator package based on the received parameters and places the validator package so that the notary can find it.

%Pseudo-code for smart contract functionality
\begin{algorithm}
\caption{Request validation (smart contract functionality)}
\label{algo:req_validation}
\begin{algorithmic}[1]
\Procedure{request\_validation(}{}
    \State $\qquad notary\_id, data\_hash, data\_url, data\_pw,$
    \State $\qquad encrypted, secret, new\_secret\_hash)$
    \State \# Abort execution if data is not certified
    \State \textbf{assert} $self.is\_certified(data\_hash)$
    \State $self.cert[data\_hash].used\_code \gets used\_code$
    \State \# Abort execution if wrong secret is provided
    \State \textbf{assert} $check\_secret(data\_hash, secret)$
    \State \# Renew secret hash
    \State $self.cert[data\_hash].secret\_hash \gets new\_secret\_hash$
    \State \# Creating validation package
    \State $vp \gets new validation\_package(data\_hash,$
    \State $\qquad notary\_id, data\_url, data\_pw, encrypted)$
%    \State $self.cert.vp.data\_hash \gets data\_hash$
%    \State $self.cert.vp.notary \gets notary\_id$
%    \State $self.cert.vp.data\_url \gets data\_url$
%    \State $self.cert.vp.data\_pw \gets data\_pw$
%    \State $self.cert.vp.encrypted \gets encrypted$
    \State \# Make validation package available for notary
    \State $self.place\_vp(notary\_id, data\_hash, vp)$
    \State \Return true
\EndProcedure
\end{algorithmic}
\end{algorithm}

On acceptance, the notary calls the smart contract procedure \verb!accept_vp! in Listing \ref{algo:sc:accept_vp}. First, the moving smart contract makes sure that the caller is the notary specified in the validator package. It would also reveal if the certificate were non-existent or no validator package had been created. Afterward, it returns the code specified in the validator package.

%Pseudo-code for smart contract functionality
\begin{algorithm}
\caption{Accept validation package (smart contract functionality)}
\label{algo:sc:accept_vp}
\begin{algorithmic}[1]
\Procedure{accept\_vp($data\_hash$)}{}
    \State $sender \gets self.read\_sender\_address()$
    \State \# Abort execution if sender is not the requested...
    \State \# ... notary or certificate does not exist
    \State \textbf{assert} $is\_requested\_notary(data\_hash, sender)$
    \State $self.cert[data\_hash].vp.accepted \gets$ true
    \State $code\_id \gets self.cert.[data\_hash].used\_code$
    \State $code \gets self.get\_code(code\_id)$
    \State \Return $code$
\EndProcedure
\end{algorithmic}
\end{algorithm}

Finally, after the client created a certificate and further a validator package directed toward a notary, and after the notary executed that validator package, the moving smart contract has to evaluate the validation. The notary calls the smart contract procedure shown in Listing \ref{algo:sc:validate}. Again, this procedure starts with an assertion to assure that the call from the notary is correct and permitted. The moving smart contract now compares the new result (i.e., hash) from the notary with the stored (i.e., calculated by the client) result in the certificate. On a match, the validator package of that certificate is tagged as validated, and a timestamp is added.

%Pseudo-code for smart contract functionality
\begin{algorithm}
\caption{Validate data (smart contract functionality)}
\label{algo:sc:validate}
\begin{algorithmic}[1]
\Procedure{send\_notary\_result($data\_hash, n\_result$)}{}
    \State $sender \gets self.read\_sender\_address()$
    \State \# Abort execution if sender is not the requested...
    \State \# ... notary or certificate does not exist
    \State \textbf{assert} $is\_requested\_notary(data\_hash, sender)$
    \State \# Check if results (hashes) match
    \State $match \gets (data\_hash = n\_result)$
    \If{$match$}
        \State $self.cert[data\_hash].vp.validated \gets$ true
        \State $self.cert[data\_hash].vp.timestamp \gets self.time()$
    \EndIf
    \State \Return $match$
\EndProcedure
\end{algorithmic}
\end{algorithm}

\subsection{Implementation Results}

% Observations made during implementation and testing of the above code. 

To test the concept, the presented implementation needs to be deployed on a running system. Therefore, the developed moving smart contract is deployed on the Bloxberg blockchain, while the client-side code can be run on any personal computer. The server-side tool has to be deployed on a Bloxberg node. For this, the node hosted by the RTC Digital Business Processes at the University of Applied Sciences Hamburg was available. This section summarizes the results that could be observed during implementation and test runs.

% Describe the test run (how was data supplied, which software was used)

The test run was successful, as the client used the software stored in the moving smart contract and added a hash of arbitrary data to Bloxberg. The client could also invoke validation of the hash by the notary via the moving smart contract. Finally, the comparison performed by the moving smart contract reported a match.

Since data privacy is a concern in this project, it was necessary to use encryption in order to prevent access to the data via the information posted on the blockchain. The availability of public keys of the blockchain nodes is suitable here. 

The ability to deploy and call smart contracts is a necessity to implement the presented concept. On Bloxberg, everyone can access the (public) blockchain, add and call new smart contracts via an open accessible node (hosted by the Max Planck Digital Library). Also, the internal Bloxberg crypto-currency ``Berg'', needed for transaction fees, is available via a faucet at no cost. In the presented case, it turned out to be an advantage to let clients use the open accessible node instead of their notary to interact with the moving smart contract. Otherwise, a malicious notary could misuse its power of being both, a notary and access provider at the same time.

The test runs also revealed certain limitations of the concept. 
Foremost, the process of creating executable scripts based on  source code stored on the blockchain turned out to be error-prone. Since the source code is handled as a string, little errors such as missing line breaks can cause syntax violations and, finally, compilation errors. This would be hard to control in a productive system.
Also, the execution environment, including the interpreter and available libraries, needs to offer all necessities to make the scripts properly executable.  
The source code must also be written following defined rules in order to make the resulting scripts operable by the client/notary tool. This includes rules defining the parameters for function calls as well as return values.  
% Foremost, the process of code compilation turned out to be error-prone. Storing Python code on the blockchain seemed efficient, but tests with different setups and software eventually yield compilation errors, which would be hard to control in a productive system.

Furthermore, the current state of the implementation lacks processes to add new notary nodes or software/source code. Trustable notaries and reviewed software build the foundation of the presented concept. Therefore, further implementations need to include sophisticated governance concepts, controlling the availability of both. Such concepts can be part of future research works.

%-- Blockchain as a communication layer has its downsides -> sesitive data must be encrypted. Public keys must therefore be available. -- SC functionalities must always be triggered. -> SC can't "push" data, user always trigger an function and get a result. -- Client and notary are controlling each other, but this doesn't work, if notary is providing the blockchain access. -- No concept yet implemented, how to prevent spam attacks from clients (Bergs are for now for free) -- Implementation lacks of a sophisticated government layer/model, ensuring code and notary integrity. -- Execution environment can be a problem for result reproducebility, docker container could help here --> yml-files could be stored on-chain. -- When a client loses his/her secret, data provenance can't be proven, also no way to return secret --> better approach to have the client's register somehow? -- Spam attacks on notary can happen -- Using on-chain code, creating runable scripts out of it --> works --> desired functionality is given -- no incentive system for notary implemented yet -- code/software must be created following a certain schema in order to make it interoperable with the client/notary tool. -- How can the client/notary tool be recieved and checked for integrity? -- recieving and creating runable scripts results in extra computation, future approach --> create script once and then only  check version/hash of script by comparing the version number/hash with on-chain?
\section{Evaluation and Discussion}
\label{sec:evaluation}

In the system described in this paper, there are four elements: (1) parties (entities that can request and execute software and retrieve and input data to that software), (2) software, (3) data, and calculation (4) results. The calculation result depends on the used software and data. The scenario of research data certification defines the software so that it delivers the same result for the same input data, but it should deliver a different result if the input data differs. This definition describes a robust hash function.

The central goal of moving smart contract is to distribute the software to the parties (clients, notaries, and peers) to make sure that every party can handle given data in the same way. Combined, this implies that parties receiving their software via the moving smart contract deliver the same result when processing the same input data. The implementation results have confirmed this behavior.

There are two possibilities for deviating outcomes when two parties are processing data with software provided by the moving smart contract: (1) the two parties used different software, or (2) the two parties used different input data. Since the presented implementation does not control execution, if the used software is the same as the one transferred, they can cheat. 
% For the discussion, take into account the previously defined roles of client and notary. 
However, a client has little incentive to cheat. If she decides to do so, her results will probably never be validated. So while having the possibility to introduce a manipulated state into the system, it will not be possible to validate the result by a notary, i.e.\ the added state will eventually become useless. Consequently, the only reason to add such information to the blockchain would be to block a notary. Therefore, in the future, one needs to take measures to avoid such behavior.

The concept, even more, limits the incentive for notaries to cheat. In principle, they can report any result back to the moving smart contract. They can validate a result without checking it; in this case, at the latest, after complete publication (i.e., disclosure of the processed data by the client), any party will be able to (in-) validate the result, thereby proving the notary's misbehavior. They can also choose to invalidate a result on purpose. In this case, the client could choose to publish her data or let another notary validate her result. Even this could be proof of notary misbehavior. For notaries, proven misconduct can have a severe effect. This paper suggests moving smart contracts, especially for consortium blockchains. Misbehavior could lead to a ban from the consortium. For research institutions, as in the case of the Bloxberg blockchain, such a ban could result in a very negative impact on the institution's reputation. 

Similarly, parties could cheat regarding the used data. The results would be similar. However, a client could deliberately share incorrect data with a notary and the correct data with another notary to collect proof for the misbehavior of a particular notary. There are possible solutions to this problem (e.g., parties could be required to do a handshake on data). Additionally, one could argue that there is no motivation for a notary to fail a validation; therefore, such cases should be irrelevant for judging a notary. Finally, this paper aims to present a concept that introduces a trusted state onto the blockchain based on off-chain data. If a client decides to misuse the system to undermine the credibility of a notary, it first undermines the credibility of her data.

The discussion above does not include errors resulting from accidents. Data could get corrupted during transportation, a client could advertise the wrong software, or a client could forget about the secret she stored together with her result (see implementation details). Discussing such problems will remain part of future research and development and is out of the scope of the present work.

\section{Conclusions}
\label{sec:conclusion}

This paper presented and evaluated the concept of moving smart contracts. The basic idea behind this concept is to deploy a smart contract that administers software available via a blockchain. When a new state on the blockchain is based on off-chain data, involved parties can always rely on the moving smart contract to provide the same software (guaranteed by its availability on the blockchain) to process that data. In this way, the initial processing of the off-chain data can be repeated and validated by other parties that gain access to the original data---the introduced state on the blockchain gains trust. Furthermore, the notary is introduced. A notary is a node of the blockchain network selected by the client, introducing the off-chain data. The concept of the moving smart contracts allows the notary to validate the client's data processing result and, as such, the state introduced on the blockchain.

Research data certification proved to be a good use case to evaluate the concept. In this, research data is hashed and the hash is stored on the blockchain to prove data provenance later on. Every party (re-)producing the hash must use precisely the same algorithm to create the hash value. If every party uses the same piece of software, this is guaranteed. Additionally, data privacy can be a concern early on in the research publication process. This paper presented the implementation of research data certification using moving smart contracts. The moving smart contract administered software for hashing and spread this to the client and notary in the implementation. The parties successfully performed certification and validation.

While there is undoubtedly future research and development work left (e.g.\
the processes to add and review new software and notaries need to be developed, the way to store and execute the software needs to be more robust) the present implementation convinced that the concept of moving smart contracts is feasible and adds a new way to introduce trusted states based on off-chain data onto the blockchain.

\section*{Acknowledgment}

The work was funded by the German Federal Ministry of Education and Research (BMBF)
under the grant marks 16QK03A and 16QK03C. The responsibility for the content of
this publication lies with the authors.

\ifCLASSOPTIONcaptionsoff
  \newpage
\fi

\bibliographystyle{IEEEtran}
\bibliography{IEEEabrv,msc.bib}

\end{document}